\documentclass{aa}

\usepackage[utf8]{inputenc}
\usepackage{color}
\usepackage{ragged2e}
\usepackage{siunitx}
\usepackage{lmodern}

\usepackage{graphicx}
\usepackage[varg]{txfonts}

\usepackage{graphicx}
\usepackage{amsmath}
\usepackage{amssymb}
\usepackage{dcolumn}
\usepackage{xcolor}
\usepackage{bm}
\usepackage{lineno}
\usepackage{mathtools}
\usepackage{ulem}

\begin{document}

\title{Sudden depletion of Alfvénic turbulence in the rarefaction region of corotating solar wind high speed streams at 1 AU: possible solar origin ? }

%\titlerunning{Turbulence features in high speed streams}
\titlerunning{Turbulence features in fast wind streams}
\authorrunning{Carnevale et al.}

%\correspondingauthor{Roberto Bruno}
%\email{roberto.bruno@inaf.it}

\author{G. Carnevale\inst{1,3} \and R. Bruno\inst{2} \and R. Marino\inst{3} \and E. Pietropaolo\inst{1} \and J.M. Raines\inst{4}}

\institute{University of L'Aquila, Dept. of Physical and Chemical Sciences, Via Vetoio 48, 67100 Coppito (AQ) , Italy
\and
INAF-Institute for Space Astrophysics and Planetology, Via del Fosso del Cavaliere 100, 00133 Rome, Italy
\and
Laboratoire de M\'ecanique des Fluides et d\'Acoustique, CNRS, \'Ecole Centrale de Lyon, Universit\'e Claude Bernard Lyon 1, INSA de Lyon, F-69134 \'Ecully, France
\and
University of Michigan, Dept. of Climate and Space Sciences and Engineering, 
Ann Arbor[MI], United States}

%\date{Received ; accepted }
\abstract{A canonical description of a corotating solar wind high speed stream, in terms of velocity profile, would indicate three main regions: a stream interface or corotating interaction region characterized by a rapid flow speed increase and by compressive phenomena due to dynamical interaction between the fast wind flow and the slower ambient plasma; a fast wind plateau characterized by weak compressive phenomena and large amplitude fluctuations with a dominant Alfv\'enic character; a rarefaction region characterized by a decreasing trend of the flow speed and wind fluctuations dramatically reduced in amplitude and Alfv\'enic character, followed by the slow ambient wind. Interesting enough, in some cases the region where the severe reduction of these fluctuations takes place is remarkably short in time, of the order of minutes, and located at the flow velocity knee separating the fast wind plateau from the rarefaction region. The aim of this work is to investigate which are the physical mechanisms that might be at the origin of this phenomenon. We firstly looked for the presence of any tangential discontinuity which might inhibit  the propagation of Alfvénic fluctuations from fast wind region to rarefaction region. The absence of a clear evidence for the presence  of this discontinuity between these two regions led us to proceed with ion composition analysis for the corresponding solar wind: we analysed minor ions parameters as tracers of the source regions looking for any abrupt variation which might be linked to the phenomenon observed in the wind fluctuations. In the lack of a positive feedback from this analysis, we finally propose a mechanism related to the magnetic field topology at the source region of the stream. We invoke a mechanism based on interchange reconnection experienced by the field lines at the base of the corona, within the region separating the open field lines of the coronal hole, source of the fast wind, from the surrounding regions mainly characterized by closed field lines. 
Obviously, a further possibility is that the observed phenomenon might be due to the turbulent evolution of the fluctuations during the expansion of the wind. However, we believe that this mechanism would not explain the sudden difference observed in the character of the fluctuations on either side of the velocity knee.
In addition, the degree of the observed reduction of the amplitude and Alfvénicity of the fluctuations appears to be related to the inclination of the heliomagnetic equator with respect to the ecliptic plane, being more dramatic for larger tilt angles. This kind of study will greatly benefit from the Solar Orbiter observations during the future nominal phase of the mission when it will be possible to link remote and in-situ data and from radial alignments between Parker Solar Probe and Solar Orbiter.}
\maketitle
%
%
%
%
%%%%%%%%%%%%%%%%%%%%%%%%%%%%%%%%%%%%%%%%%%%%%
\section{Introduction}
\label{sec:intro}
%%%%%%%%%%%%%%%%%%%%%%%%%%%%%%%%%%%%%%%%%%%%%
% add references
The solar wind is a plasma flow, electrically neutral, which emanates from the basis of the solar corona, permeating and shaping the whole heliosphere. During solar activity minima, when the meridional branches of the polar coronal holes (CH hereafter) reach the equatorial regions of the sun, an observer located in the ecliptic plane would record a repeated occurrence of fast (700-800 km/s) and slow (300-400 km/s) wind samples. The balance between the two classes of wind would change during the evolution of the 11-year solar cycle depending on changes in the topology of the heliomagnetic equator and its inclination on the ecliptic plane. The fast wind originates from unipolar open field line regions, typical of the coronal holes \citep{Schatten69, Hassler99}. The slow wind origin is much more uncertain, although we know this class of wind is generated within regions mainly characterized by closed field lines configuration that likely inhibit the escape of the wind \citep{Wang90, Antonucci2005, Bavassano97, Wu2000, Hick99}. Indeed, the different composition and mass flux of the slow wind and the different degree of elemental fractionation  with respect to  the corresponding photospheric regions, strongly suggest that such slow plasma flow could be initially magnetically trapped and then released \citep{Geiss95a, Geiss95b}. Interchange reconnection process plays a fundamental role in opening up part of the closed field lines linked to the convective cells at photospheric level \citep{Fisk99, Fisk2001, Schwadron2002, Schwadron2005, Wu2000, Fisk2020}. This process should preferentially develop close to the border of the CH, influencing as well the neighbour regions, i.e. the coronal hole boundary layer (hereafter CHBL), and the closed loop corona. It has been empirically proven that the expansion factor of the magnetic field line configuration is anti-correlated to the wind speed \citep{Wang90}. A smaller expansion factor, typical of the central region of the coronal hole, would therefore produce a faster wind while larger expansion factor, typical of the border of the coronal hole, would produce instead a slower wind \citep{Wang90, Wang2006, D'Amicis2015}. Several other contributions to the slow wind come from more specific processes like, e.g. the release of plasmoids by helmet streamers or the continuous leakage of plasma from the highest regions of the large coronal loops where magnetic pressure is no longer largely dominating the thermal pressure \citep[among others]{Bavassano97, Hick99, Antonucci2005}. 
During the wind expansion, magnetic field lines from coronal holes of opposite polarities are stretched into the heliosphere by fast wind streams being separated by the heliospheric current sheet, an ideal plane, magnetically neutral, permeated by slow wind plasma. During the expansion into the interplanetary space, because of solar rotation, the high speed plasma would overtake and impact on any slower plasma ahead of it creating a compression region \citep{Hundhausen1972}.
% FIGURE 1
\begin{figure}[h!]
	\centering
	\includegraphics[width=7cm]{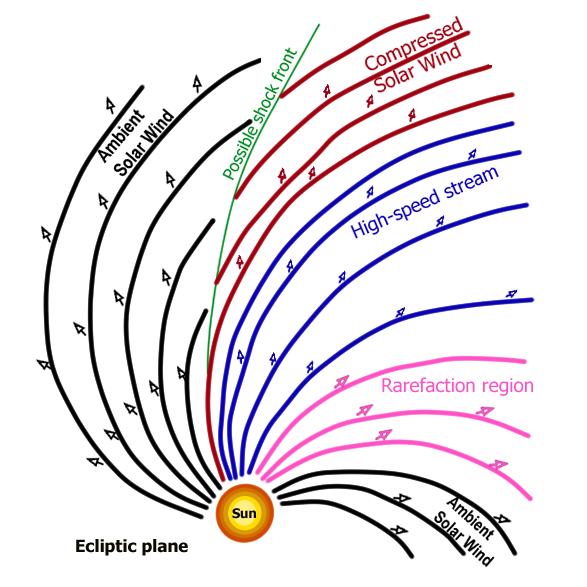}
	\caption{Sketch of a stream structure in the ecliptic plane, adapted from Hundhausen 1972. The spiral structure is a consequence of solar rotation. The spiral inclination changes as the solar wind velocity changes. When the high-speed stream (in blue) compresses the slower ambient solar wind (in black), a compression region downstream (in red) and a rarefaction region upstream (in pink) are formed.}
	\label{fig01}
\end{figure}
As a consequence of this interaction, a so called co-rotating interaction region (CIR), characterized by a rather rapid increase in the wind speed, from values typical of the slow wind (300-400 km/s) to values typical of fast wind (700-800 km/s), forms at the stream-stream interface. The CIR is characterized by strong compressive phenomena (red zone in Figure \ref{fig01}) affecting both magnetic field and plasma \citep{Richardson2018}. The CIR is followed by a region where the wind speed persist at its highest values (blue zone in Figure \ref{fig01}), resembling a sort of plateau, while the other plasma and magnetic field parameters rapidly decrease to remain rather stable across its extension. Beyond this region, highlighted at times by a a sharp knee in the radial velocity, the wind speed decreases monotonically to reach values typical of the following slow wind (black zone in Figure \ref{fig01}), which permeates the interplanetary current sheet. This decreasing wind speed region is slightly more rarefied than the high speed region which immediately follows the CIR and is commonly called rarefaction region (pink zone in Figure \ref{fig01}). The three regions described above can be considered the imprint in the interplanetary medium of the coronal structure from which the wind originated \citep{Hundhausen1972}. The fast wind plateau corresponds to the core of the CH while CIR and rarefaction region correspond to the same CHBL encircling the CH, though within the CIR the CHBL is compressed by the dynamical interaction described earlier between fast and slow wind while, within the rarefaction region the CHBL is stretched by the wind expansion into the interplanetary medium.
Beyond a few solar radii, the  solar wind (both fast and slow) becomes supersonic and super-Alfv\'enic. Fluctuations of interplanetary magnetic field and plasma parameters start to develop their turbulent character since the first tens of solar radii \citep{Kasper2019} and, in particular, turbulence features within fast and slow wind differ dramatically \citep[and references therein]{BrunoCarbone2013}. Fast wind is rather uncompressive and mostly characterized by strong Alfv\'enic correlations between velocity and magnetic field fluctuations, though carrying also compressive fast and slow magnetosonic modes  \citep{Marsch90,Marsch93,Klein2012, Howes2012, Verscharen2019}. The origin of  Alfv\'enic and compressive fluctuations is rather different. The former appear to be generated at large scales through the shuffling of the magnetic field lines foot-points (by photospheric motion \citep{Hollweg2006}) and at small scales by magnetic reconnection processes together with fast/slow magnetosonic modes \citep{Parker57, Lazarian2015, Kigure2010, Cranmer2018}. On the other hand, following \cite{Marsch93} and \cite{Tu94} compressive modes may also originate during the wind expansion from a superposition of pressure balanced structures of solar origin and magnetoacoustic fluctuations generated by the dynamical interaction of contiguous flux-tubes. Uncompressive and compressive fluctuations eventually experience a non-linear cascade to smaller and smaller scales.
Alfv\'{e}nicity in the slow wind is in general much lower with compressive effects affecting more the plasma dynamics and reflecting the complex magnetic and plasma structure of its source regions, close to the heliomagnetic equator %\citep{Tu95,BrunoCarbone2013}. 

This paper deals with a peculiar phenomenon that we observed in the Alfv\'{e}nic turbulence when passing from the fast wind plateau to the rarefaction region. The border between these two regions  is often marked by a rather sharp knee in the velocity profile. As detailed in the following, We noticed that, while downstream the knee, fluctuations have large amplitude and are highly Alfv\'enic, in the upstream region, fluctuations dramatically reduce their amplitude and the Alfvénic correlation weakens, beginning to fluctuate between positive and negative values. Interesting enough, the region where the remarkable depletion of these fluctuations takes place can be very short in time, of the order of tens of minutes, and is located around the velocity knee.
To our knowledge, no specific analyses, before the present work, have ever been devoted to understand the reason of such a sudden change of the Alfv\'enic turbulence across the velocity knee. \par 
As we will explain in the following section, after analyzing the turbulent character of the stream, we first looked for the presence of a tangential discontinuity between the fast wind region and the rarefaction region. Then, we made a composition analysis of the minor ions of the solar wind. Finally, we suggested a mechanism that we believe may explain the rapid depletion of fluctuations and Alfvénicity observed at the beginning of the rarefaction region.
%
%
%%%%%%%%%%%%%%%%%%%%%%%%%%%%%%%%%%%%%%%%%%%%%
\section{Data analysis}
\label{sec:data}

We analyzed 3 second averages of magnetic field and plasma observed at 1AU from the WIND spacecraft and 2 hour averages minor ions parameters from the ACE spacecraft during the last minimum of solar activity. The data refers to the Heliocentric Earth Ecliptic (HEE) system, which has an X axis pointing along the Sun-Earth line, and a Z axis pointing along the ecliptic north pole \citep{Thompson2006}. In particular, after a visual inspection of several co-rotating high speed streams towards the end of solar cycle 24, we focused on a series of wind streams in which the phenomenon described above appears, though for sake of brevity here we report results related to the high speed stream observed on 3-9 August 2017, representative of those streams showing a similar fluctuations reduction phenomenon. We also show, for comparison, a case in which the reduction in fluctuations and Alfvénicity occurs more slowly, without being able to identify a clear velocity knee in this case.
Since the magnetic field and plasma data collected by WIND are not provided with synchronized time stamps, we re-sampled the time series with a 6 second cadence in order to build up a single merged data-set with a uniform time base to be used to characterize the statistics of the turbulent fluctuations of the solar wind parameters under study. 
For the minor ions analysis we were forced to employ the 2 hour cadence being the observations with the highest time resolution available during the selected streams.

To characterize the turbulence in the solar wind we built the Els\"asser variables $ \bf{z^\pm} = \bf{v} \pm \bf{b} $ \citep{Elsasser1950}, where $\bf{v}$ is the velocity vector and $\bf{b}$ is the magnetic field vector expressed in Alfv\'en units \citep{Tu95, BrunoCarbone2013}. 

The second order moments linked to Els\"asser variables are \citep{BrunoCarbone2013} :
\begin{subequations}
\begin{align}
   e^{\pm} &=\frac{1}{2} \langle \left(\bf{z^{\pm}} \right)^2 \rangle & \: (energies\; related\; to\; \mathbf{z^+} \;and\; \mathbf{z^-})\\
   e_{v}&=\frac{1}{2} \langle v^2 \rangle & (kinetic\; energy)\\
   e_{b}&=\frac{1}{2} \langle b^2 \rangle & (magnetic\; energy)\\
   e_c &= \frac{1}{2} \langle \bf{v} \cdot \bf{b} \rangle & (cross-helicity)
\end{align}
\end{subequations}
where angular brackets indicate averaging process over the established time range. In our case $e^{\pm}$, $e_{v}$, $e_{b}$ and $e_{c}$ represent the variances calculated on hourly moving window for the whole stream.
In order to describe the degree of correlation between v and b, it is convenient to use normalized quantities:\\
\begin{subequations}
\begin{align}
\sigma_c &= \frac{e^+ - e^-}{e^+ + e^-}  & \;\; (normalized \;cross-helicity)  
\label{def_sigma_c}   \\
\sigma_r &= \frac{e_v - e_b}{e_v + e_b}  & \;\; (normalized \;residual \; energy) \label{def_sigma_r}
\end{align}
\end{subequations}
Where $-1 \leq \sigma_c \leq 1 $ and $-1 \leq \sigma_r \leq 1 $.

\vspace{7mm}
% FIGURE 2
\begin{figure}
	\centering
	\includegraphics[width=9cm]{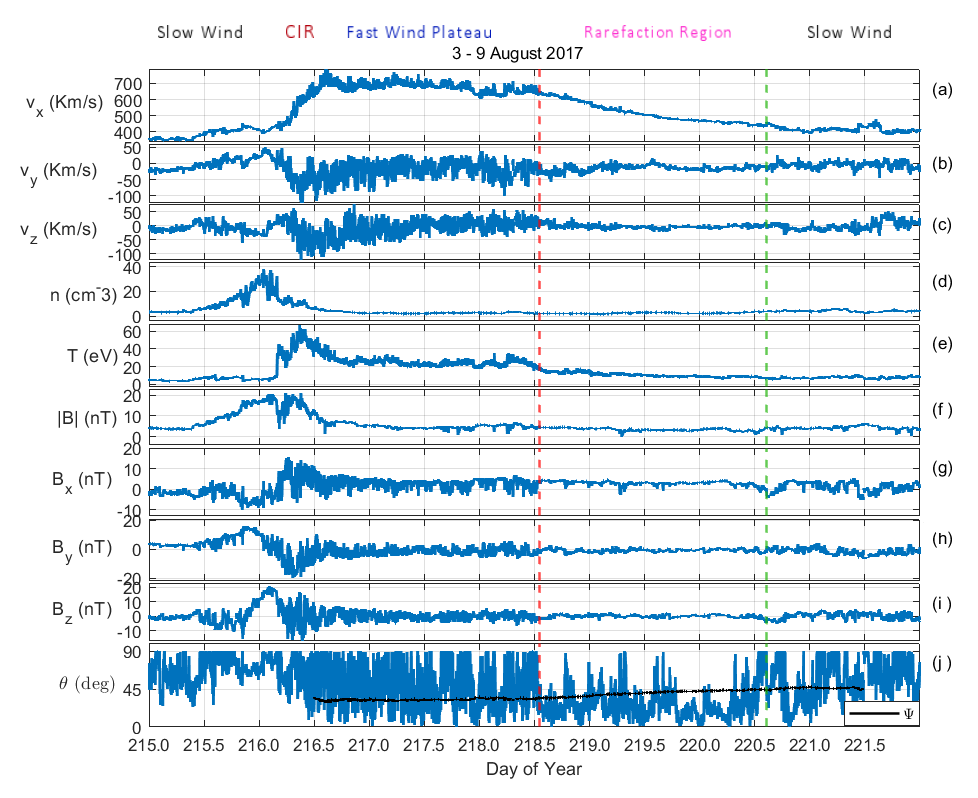}
	\caption{Temporal profiles of solar wind parameters from $3^{th}$ to $9^{th}$ August 2017, at 6 seconds time resolution. Panels show from top to bottom: proton velocity components in HEE (Heliocentric Earth Ecliptic) reference system, respectively $v_x$ (panel a), $v_y$ (panel b), $v_z$ (panel c); proton number density (d); proton temperature (e); magnetic field magnitude (f); magnetic field components in HEE, respectively $B_x$ (panel g), $B_y$ (panel h), $B_z$ (panel i); angle between the average magnetic field direction and the radial direction (j). In the latest panel the black dashed curve represent the local Parker spiral direction, referred to the high-speed region and rarefaction region. The red dashed line in each panel denotes the velocity knee between fast wind region and rarefaction region. The green dashed line in each panel denotes the location of the Tangential Discontinuity found at almost the end of the rarefaction region (see section \ref{sec:data}). On the higher part of the figure (outside the panels) are indicated the different solar wind region, vertically interpreted, with color-correspondence to sketch of Figure \ref{fig01}.}
	\label{fig02}
\end{figure}
The multiple panels of Figure \ref{fig02} show some of the plasma and magnetic field features characterizing the particular stream we chose observed by WIND on August 2017. From top to bottom we show the time profiles of the three velocity components $v_x$, $v_y$ and $v_z$ (panels a to c), proton number density (panel d), proton temperature (panel e), magnetic field intensity (panel f), magnetic field components $B_x$, $B_y$ and $B_z$ (panels g to i) and, at the bottom (panel j), the angle $ \theta = cos^{-1} \left(|B_{x}| / |B| \right) $ (limited between $0$ and $90^{\circ}$) between the local field and the radial direction, respectively. This is a canonical corotating high-speed stream in which it can be easily recognized an interaction region, roughly occurring around noon of August the $4^{th}$ and characterized by strong magnetic field intensity and plasma number density enhancements followed by a clear proton temperature increase, as expected for a compressive region, because of the dynamical interaction of this stream with the slow wind ahead of it \citep{BrunoCarbone2013}. This interaction region, within which the wind speed rapidly increases from about 400 km/s to about 700 km/s, is followed by a speed plateau about two days long. The latter is characterized by fluctuations in velocity and magnetic field components of remarkable amplitude and rather constant values of number density \begin{math} { n \sim 3 \; cm^{-3} } \end{math} and field intensity \begin{math}{|B| \sim 5 \; nT }\end{math}. In addition, the temperature profile (panel e), although lower than that within the preceding interaction region (CIR), is much higher than in the surrounding low speed regions. The high speed plateau is then followed by a slowly decreasing wind speed profile identifying the rarefaction region of this corotating stream. These features are typical for most of the high speed corotating streams we examined. What is less common is the abrupt depletion of large amplitude fluctuations detected in the high speed plateau, across the velocity knee (defined later, when describing Figure \ref{figura5}) indicated in Figure \ref{fig02} by the vertical dashed red line around day 218 (6 August) at noon. This phenomenon is highlighted by the profile of the curve in the bottom panel j related to the angle $\theta$ between the magnetic field vector and the radial direction. The difference in the behavior of this parameter before and after the velocity knee is dramatic and, as shown later on, this transition is rather fast, of the order of minutes.
Large angular fluctuations between the magnetic field vector and the radial direction are due to the presence of large amplitude, uncompressive  Alfv\'{e}nic fluctuations populating the high speed plateau that force the tip of the magnetic field vector to randomly move on the surface of a hemisphere centered around the background mean field direction \citep[e.g. see figure 4 in ][]{Bruno2001}.
% FIGURE 3
\begin{figure}
	\centering
	\includegraphics[width=9cm]{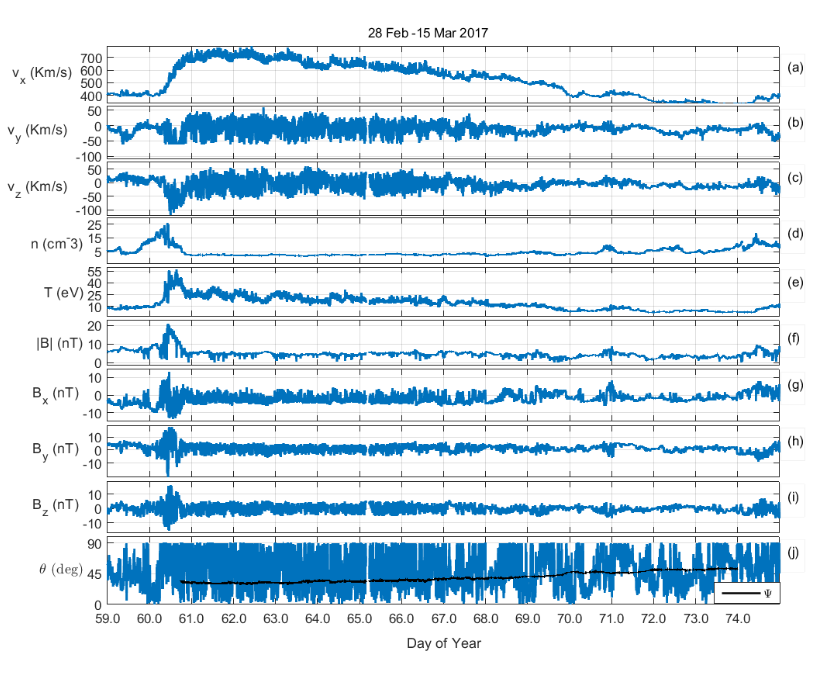}
	\caption{Temporal profiles of solar wind parameters from $28^{th}$ February to $15^{th}$ March 2017, at 6 seconds time resolution. Panels show from top to bottom: proton velocity components in HEE (Heliocentric Earth Ecliptic) reference system, respectively $v_x$ (panel a), $v_y$ (panel b), $v_z$ (panel c); proton number density (d); proton temperature (e); magnetic field magnitude (f); magnetic field components in HEE, respectively $B_x$ (panel g), $B_y$ (panel h), $B_z$ (panel i); angle between the average magnetic field direction and the radial direction (j). In the latest panel the black dashed curve represent the local Parker spiral direction referred to the high-speed region and rarefaction region.}
	\label{fig03}
\end{figure}

On the other hand, not all streams have the same abrupt depletion of fluctuations after the fast wind plateau; an example is the stream of March 2017 shown in Figure \ref{fig03}. In the multiple panels there are: velocity components $v_x$, $v_y$ and $v_z$ (panels a to c), proton density (panel d), proton temperature (panel e), magnetic field magnitude (panel f), magnetic field components $B_x$, $B_y$ and $B_z$ (panels g to i) and the angle $\theta$ (panel j) between the local field and the radial direction. We can clearly identify the interaction region characterized by a compression (enhancement of proton density in panel d), followed by an increase in magnetic field, proton temperature and velocity of the solar wind, typical of high speed corotating streams.
Immediately after, there are wide fluctuations in velocity and magnetic field components in correspondence of the fast wind plateau of this stream, that has a longer duration with respect to the previous stream. In this case, unlike the previous one, even though there is a decrease in velocity and magnetic fluctuations, it occurs gradually over time (days). This gradual decreasing trend is also present in the temperature profile (panel e). Moreover, the angular fluctuations between the local field and the radial direction (panel j) change rapidly over time, without having a clear trend in the rarefaction region, unlike the previous stream. Hence the magnetic field vector fluctuates continuously around the radial direction for almost the entire duration of this second stream.

Comparing the time profiles of the angle between the average direction of the magnetic field with respect to the radial direction (panels h of Figures \ref{fig02} and \ref{fig03}), we note that in correspondence of the fast wind region of both August (Figure \ref{fig02} h) and March (Figure \ref{fig03} h) streams, most of the values are above the black solid curve, which indicates the predicted Parker spiral angle based on the wind speed, computed as $\Psi = \arctan(\Omega R / |v|)$ where R is the distance from the Sun to the Earth and $\Omega$ is the sidereal angular velocity of the Sun at the equator, i.e. $2.97 \cdot 10^{-6} rad/s$.
As a consequence, before the velocity knee, the background magnetic field is over-wound \citep{BrunoBavassano1997}, most likely as a consequence of the dynamical interaction of the fast stream with the slow wind ahead \citep{Schwenn90}. Within the rarefaction region, the time profile of this angle for the August stream shows a smaller variability with an average direction closer to the radial one and smaller than the local Parker spiral direction, as expected for a rarefaction region \citep{SchwadronMcComas2005}. This region maps in the interplanetary space the coronal hole boundary layer (CHBL) encircling the fast wind source region at the sun \citep{McComas2003}. The solar wind observed in space and emanating from this source region, because of the solar rotation, is characterized by compressive phenomena when detected ahead of the high velocity stream and rarefaction phenomena when it is observed following the high speed plateau.  The existence of this sub-Parker spiral orientation of the background magnetic field depends on the rate of the motion of open magnetic field foot points across the coronal hole boundary at photospheric level \citep{SchwadronMcComas2005}. Such motion is responsible for connecting and stretching magnetic field lines from fast and slow wind sources across the CHBL, as explained in the model by \citep{SchwadronMcComas2005}.

It is interesting to remark how different are the profiles of $\theta$ for these two streams within the rarefaction regions. For the stream of August, the $B_x$ component is generally positive and the profile of $\theta$ is largely confined to values below $45 ^{\circ} $ with sporadic jumps to larger values.  For the March stream, the $B_x$ profile is much more structured and  $\theta$ continuously fluctuates between $0^{\circ}$ and $90^{\circ}$. Moreover, for this last stream, $B_x$ continuously jumps from positive to negative values suggesting that we must be close to the heliomagnetic equator.
% FIGURE 4
\begin{figure}
	\centering
	\includegraphics[width=9.2cm]{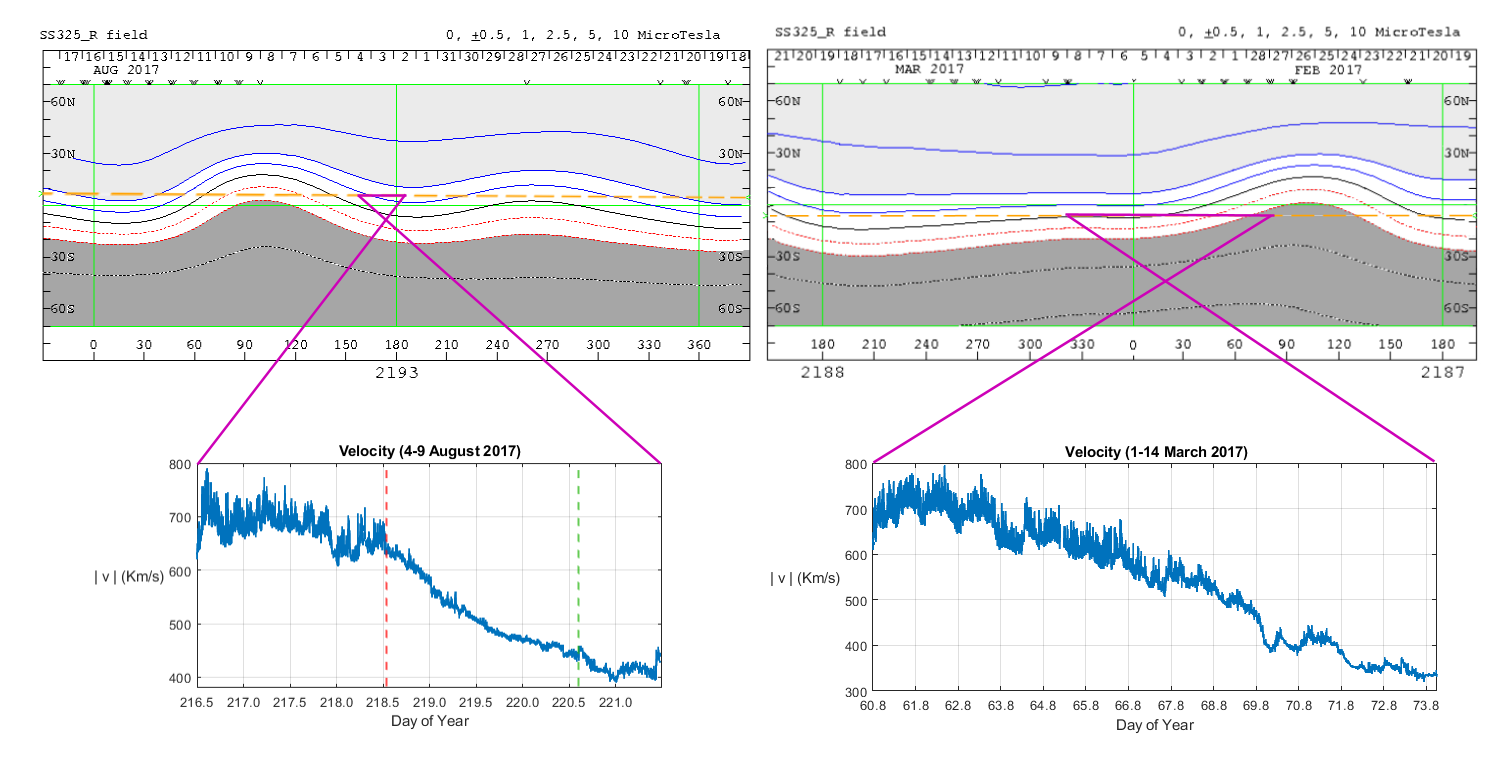}
	\caption{Source Surface Synoptic Charts at  $3.25 R_s$ related to July-August 2017 on the left side and February-March 2017 on the right side (Source: Wilcox Solar Observatory).The level curves indicate values of constant magnetic field (at $3.25 R_s$), the black line represents the heliomagnetic equator (current sheet), the two different shades of gray indicate the magnetic field polarity: dark gray and red lines correspond to negative polarity while light gray and blue lines correspond to positive polarity. The orange dashed line represent the Earth trajectory projected to the Solar surface at $3.25 R_s$ . The different purple lines correspond to the periods of the studied streams. In the bottom two panels are shown the velocity profiles of August 2017 stream (on the left) and March 2017 stream (on the right).}
	\label{figura4}
\end{figure}
As a matter of fact, in our analysis we have observed many cases like the one (August 2017) described, on the other hand we also observed several other cases where the disappearance of the Alfv\'enic fluctuations is not characterized by a sudden event like the one observed here on August the $6^{th}$. Individual streams can be different from each other and one cannot simply infer their turbulence properties without proper analysis.
As observed by \cite{Shi2021}, structures such as heliospheric current sheets can play an important role in modifying the properties of turbulence in the solar wind, also supported by 2D MHD simulations of Alfvénic waves on a current sheet \citep{Malara1996}, which showed that initially high values of $\sigma_c$ are quickly destroyed near a current sheet.
We observed that the sudden depletion of the Alfv\'{e}nic fluctuations beyond the velocity knee might be related to the pitch angle with which an observer, in the ecliptic plane, crosses the heliomagnetic equator. 
In the upper panel of the left-hand-side of Figure \ref{figura4} we show the Source Surface Synoptic Chart relative to July-August 2017 taken from Wilcox Solar Observatory. The level curves indicate values of constant magnetic field (at 3:25Rs), the black line represents the heliomagnetic equator, which runs in the middle of the heliospheric current sheet. The two different shades of gray indicate the magnetic field polarity: dark gray and red lines correspond to negative polarity while light gray and blue lines correspond to positive polarity. The orange dashed line represent the Earth trajectory projected to the Solar surface at 3:25Rs. The purple lines indicate the beginning and the end of the high speed stream studied in this work, whose speed profile is shown in the lower panel. In this case (4-8 August 2017), the heliospheric current sheet is highly inclined with respect to the trajectory of the observer, i.e. WIND s/c. In the right-hand-side of this Figure, we show, in the same format as of the left-hand-side, one of those cases for which we do not see a clear and abrupt depletion of the fluctuations. This is a high speed stream observed on March 2017. In this case, the speed profile shows a less clear velocity knee and a slow and progressive disappearance of the fluctuations without any abrupt event like the one observed in the previous stream. The main difference, in this case, is that the heliospheric current sheet is very flat and the observer experiences a sort of surfing along this structure instead of a clear and rapid crossing. 
% FIGURE 5
\begin{figure}
	\centering
	\includegraphics[width=9cm]{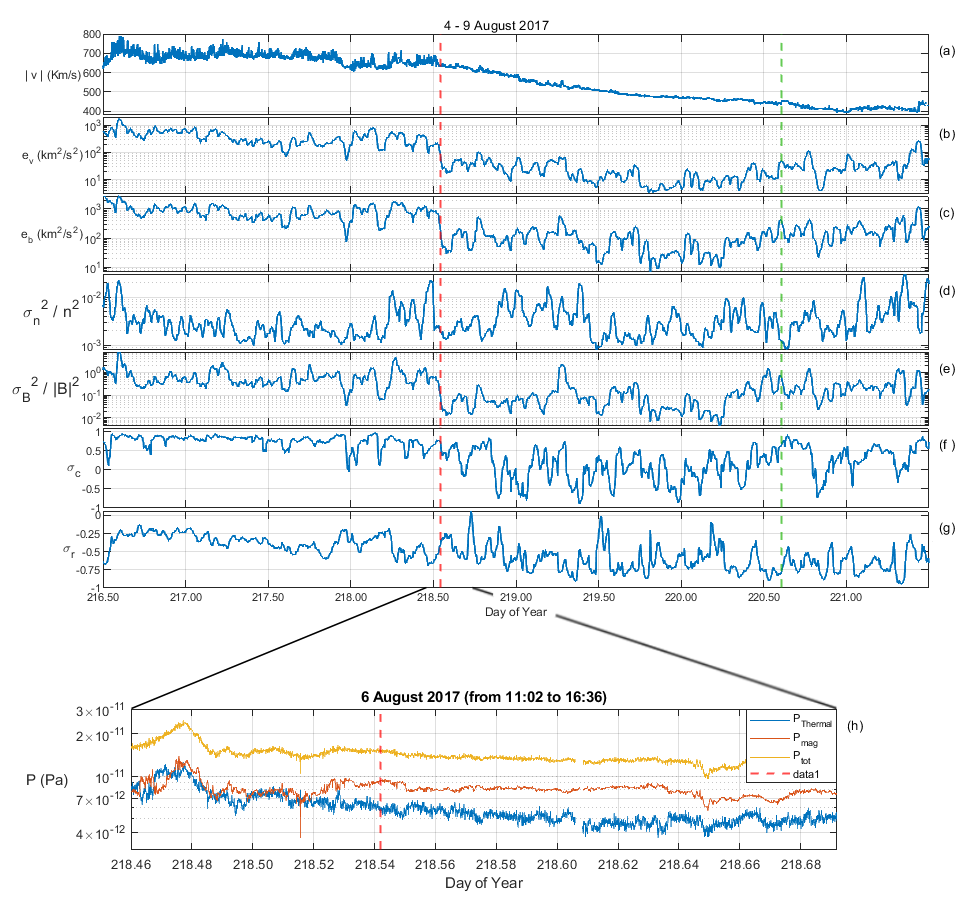}
	\caption{Fast Wind region and rarefaction region of the August 2017 stream. From top to bottom: solar wind speed at 6 seconds time resolution (a), values, computed on 1 hour moving window, of kinetic energy in semi-logarithmic scale (b), magnetic energy in semi-logarithmic scale (c), density compressibility in semi-logarithmic scale (d), compressibility of the magnetic field magnitude in semi-logarithmic scale (e), normalized cross-helicity  (f) and normalized residual-energy (g).  The bottom panel (h) shows an enlargement of the total pressure (yellow), magnetic pressure(orange) and thermal pressure (blue) trend in a small region of few hours close to the velocity knee, the latter represented by the red dashed line in each panel. The green dashed line, in panels from (a) to (g), denotes the location of the tangential discontinuity (TD) founded at almost the end of the rarefaction region (see section \ref{sec:data}).}
	\label{figura5}
\end{figure}

% Descrizione figura 5 (analisi turbolenza Agosto)
As anticipated, the high speed plateau of the stream under study is characterized by large amplitude Alfv\'{e}nic fluctuations as shown in Figure \ref{figura5}. The top panel (a) shows part of the speed profile of our corotating high speed stream. The location of the velocity knee, beyond which large amplitude fluctuations seems to turn off, is indicated by the vertical dashed red line and determined as follows. Panels b and c show hourly values of kinetic and magnetic energy, respectively, on a semi-logarithmic vertical scale.
%total variances taken from the trace of the corresponding variance matrix. 
The sharp variation in these two parameters, at the end of the speed plateau, identifies what we have defined as velocity knee. They highlight remarkably the different physical situation before and after the velocity knee and the abrupt change across it. We notice that %velocity variance 
kinetic energy is the parameter that experiences the largest decrease, of a factor $\sim 20$. As a consequence, fluctuations within this region become strongly magnetically dominated. In addition, this decrease happens in correspondence of a strong temperature reduction as shown in the previous Figure \ref{fig02} by the thermal energy parameter. This strong depletion of proton temperature reflects the thermal pressure decrease (Figure \ref{figura5} h) that starts at the velocity knee together with a corresponding increase of the magnetic pressure such to maintain the wind in a pressure balanced status, achieved at the edge of the Alfv\'{e}nic surface during the wind initial expansion \citep{SchwadronMcComas2005}. Figure \ref{figura5} (d) shows the density compressibility over time in a semi-logarithmic vertical scale, computed as ${\sigma_{n}}^2 / n^{2}$; its value remains quite low (on average $3 \cdot 10^{-3}$) both during the fast wind region and the rarefaction region, indicative of a fairly incompressible wind, given that the density remains approximately constant in both regions.
On the other hand, the compressibility of magnetic field magnitude, computed as ${\sigma_{B}}^2 / |B|^{2}$ and shown in panel \ref{figura5} (e) on a semi-logarithmic vertical scale, denotes that magnetic fluctuations, in modulus, are more compressive in correspondence of the fast wind plateau rather than in the rarefaction region; there is a sudden reduction at the velocity knee of about an order of magnitude. The two other panels, f and g, highlight the Alfv\'{e}nic nature of these fluctuations within the high speed plateau. Panel f shows high values of the normalized cross helicity $\sigma_C$ while panel g shows that the normalized residual energy, although dominated by magnetic energy, is not far from the equipartition expected for Alfv\'enic fluctuations. Beyond the velocity knee, these two parameters experience a considerable change towards a much lower Alfv\'enicity although the transition is not as abrupt as the one relative to the amplitude of these fluctuations, as discussed above. \\
The behaviour of $\sigma_C$ is similar to that recently observed in the SW from Parker Solar Probe near perihelions of the orbits $(R \le 0.3 - 0.4 \;AU) $, as described by \cite{Shi2021}, where $\sigma_c$ is usually close to $1$, implying a status of dominating outward-propagating Alfvén waves. There are also periods when $\sigma_C$ oscillates and becomes negative; these periods correspond to Parker Solar Probe observing heliospheric large-scale inhomogeneous structures, such as the heliospheric current sheet.\\
% FIGURE 6
\begin{figure}
	\centering
	\includegraphics[width=9.4cm]{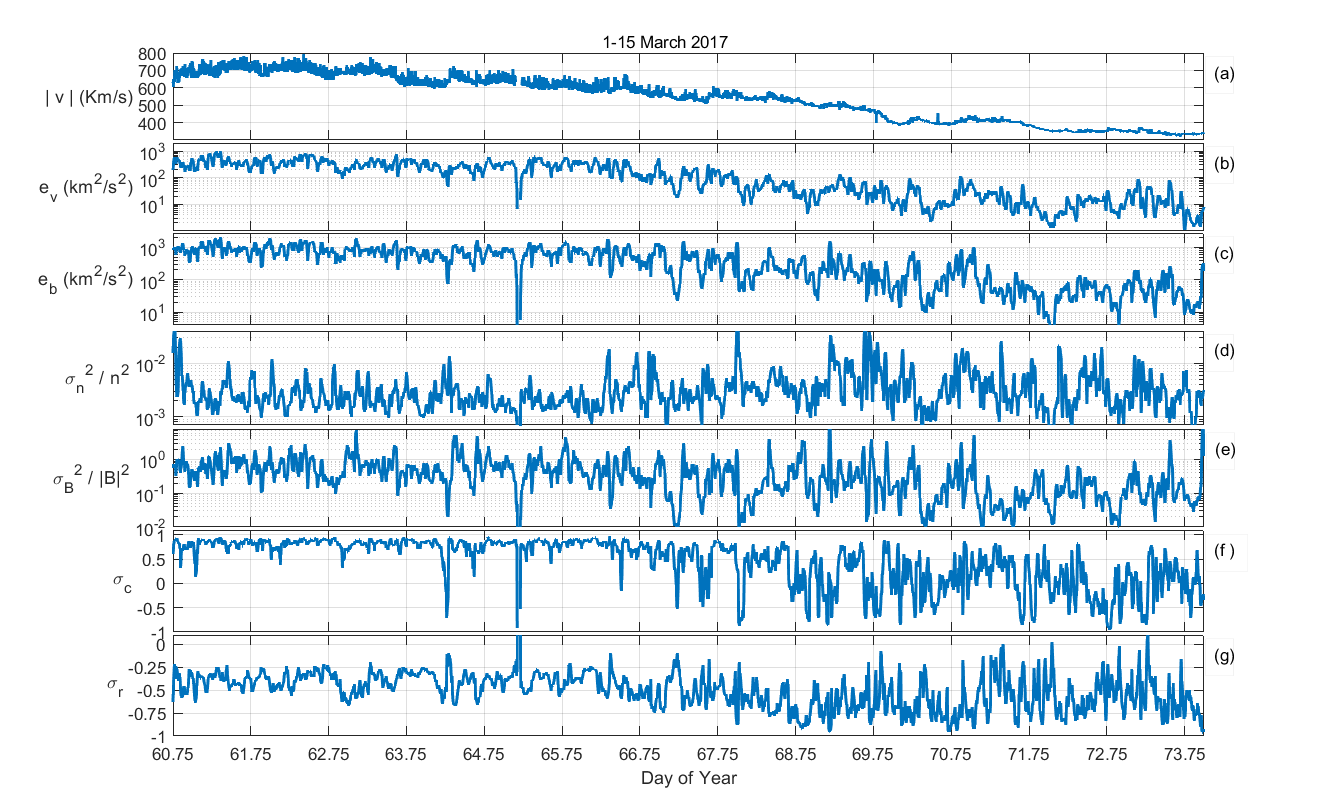}
	\caption{Fast Wind region and rarefaction region of the March 2017 stream. From top to bottom: solar wind speed at 6 seconds time resolution (a), values computed on 1 hour moving window, of kinetic energy in semi-logarithmic scale (b), magnetic energy in semi-logarithmic scale (c), density compressibility in semi-logarithmic scale (d), compressibility of the magnetic field magnitude in semi-logarithmic scale (e), normalized cross-helicity (f) and normalized residual-energy (g).}
	\label{figura6}
\end{figure}
% Descrizione figura 6 (analisi turbolenza Marzo)
The same analysis of Figure \ref{figura5} for the August 2017 stream was also carried out for the March 2017 stream and shown in Figure \ref{figura6}. In this case the decrease in fluctuations and Alfvénicity occurs very gradually over time (few days). The velocity profile (panel a) of the March 2017 stream does not allow us to define a velocity knee in this case. Furthermore, the magnetic energy (panel b) and kinetic energy (panel c) decrease more slowly in the rarefaction region of the stream, with respect to the previous one. The Alfvénic character of the fast wind remains evident, as shown by the profile of $\sigma_c$ in panel e and $\sigma_r$ in panel f. Even in this case the density compressibility (panel d) oscillates around very low values, of the order of $10^{-3}$, while in this stream the compressibility of magnetic field magnitude (panel e) does not show an effective decrease after the velocity plateau, unlike the previous case (August 2017 stream).\\
The MHD simulations of an ensemble of Alfvénic fluctuations propagating in an expanding solar wind including the presence of fast and slow solar wind streams, made by \cite{Shi2020}, showed that the decrease in $\sigma_c$ is more significant in the compression and rarefaction regions of the stream (as we observed in Figure \ref{figura5}) than within the fast and slow streams. In particular, in the rarefaction region of such simulations $\sigma_c$ rapidly decreases from 1 to 0.6 at $ R \approx 80 R_s $ and remains around this value until the end of the simulation. While leaving open the hypothesis that the phenomenon under study may develop during the radial expansion of the solar wind, the very rapid (few minutes) reduction of the fluctuations and Alfvénicity between the fast wind plateau and the rarefaction region makes us lean towards the hypothesis that this phenomenon should originate at the source regions of the solar wind. \\
The sudden changes in some of the parameters described above for the August stream, between the end of the fast wind plateau and the beginning of the rarefaction region, suggested to look for magnetic structures like tangential discontinuities (TD, hereafter), which might represent a barrier for the propagation of the Alfv\'enic fluctuations. The presence of a TD, where by definition there is no magnetic component normal to the discontinuity surface, could explain the depletion of Alfvén waves, that propagate along the magnetic field direction \citep{Hundhausen1972}. In fact, these fluctuations would not propagate across this kind of discontinuity since the magnetic component perpendicular to the discontinuity plane vanishes. This would prevent their propagation into the CHBL or rarefaction region and would produce the observed sudden depletion of the Alfv\'enic fluctuations. 

%Following indications reported in McComas et al. (1998) and Schawadron et al., (2005) we carefully analysed the region around the velocity knee in search of tangential discontinuities, i.e. a coronal hole discontinuity or CHD as named by McComas et al., 1998, which encircles the original flux-tube . 
% FIGURE 7
\begin{figure}
	\centering
	\includegraphics[width=9cm]{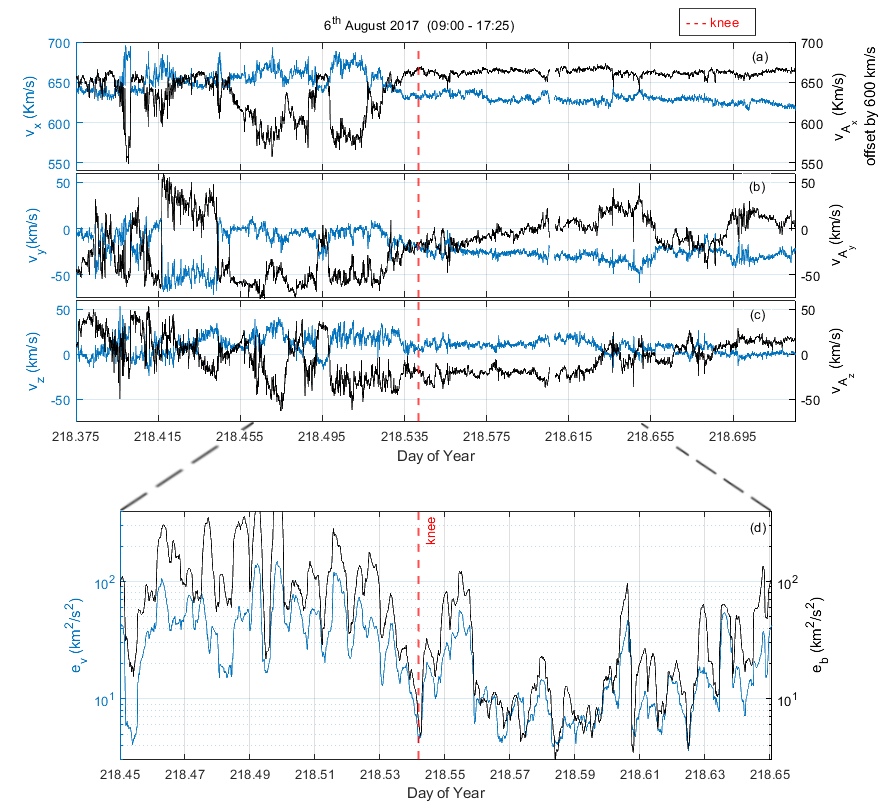}
	\caption{Top three panels: time profile of velocity components (blue lines) and magnetic field components in Alfvén units (black lines) at 6 seconds (panels a, b, c), in about an 8 hr time interval.  Bottom panel (d): 5 minute values of magnetic $e_b$ (black line) and kinetic $e_v$ (blue line) energies within a narrower time interval of almost 5 hours around the estimated velocity knee (red dashed line).}
	\label{figura7}
\end{figure}

\begin{figure}
	\centering
	\includegraphics[width=9cm]{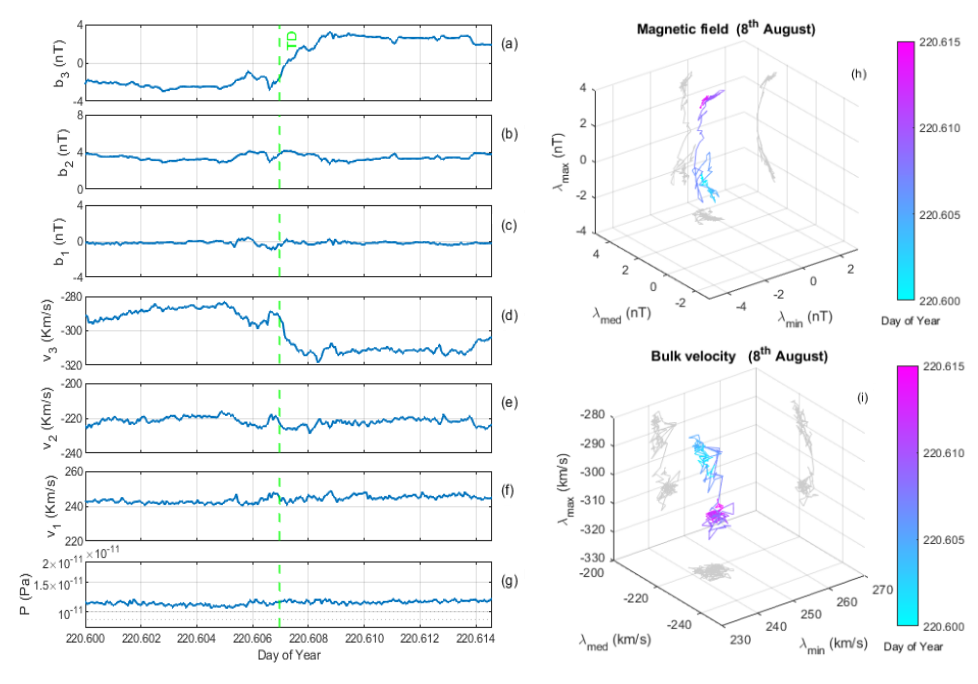}
	\caption{Left-hand-side: panels from top to bottom show the three components of magnetic field (panels (a), (b), (c))and velocity (panels (d), (e), (f)), rotated in the magnetic field minimum variance reference system, and the total pressure (panel g). The 3D figures in the right-hand-side panels show magnetic field (panel h) and velocity components (panel i) rotated in the field minimum variance reference system. All these quantities are plotted with data at 6 sec time resolution. The green dashed line, in panels from (a) to (g), denotes the location of the tangential discontinuity (TD) founded at almost the end of the rarefaction region (see section \ref{sec:data}).}
	\label{figura8}
\end{figure}

To better determine the starting time of the depletion of magnetic and kinetic fluctuations, we show in Figure \ref{figura7} the time profile of 6 sec averages of magnetic field components, expressed in Alfv\'en units, and velocity components within a short time interval, slightly longer than 8 hrs and roughly centered around the previously estimated time location of the velocity knee. There is a clear change in the amplitude of the fluctuations around the location of the dashed line. This phenomenon is highlighted by the time profile of magnetic and kinetic energies, $e_b$ and $e_v$, respectively, computed within 5 min running window, shown in the bottom panel (panel d). The sharp profile of these curves allows to better estimate the time when fluctuations are abrupt reduced, that is on day  $218^d13^h24^m$ (i.e. 6 August).
However, we did not find any relevant TD around the estimated time, which could be associated with the observed phenomenon. Thus TD can not be a valid explanation of the rapid decrease of the Alfvénic fluctuations at the velocity knee, leaving open the possibility that this phenomenon could be due to a different evolution of the solar wind during its expansion or to a possible reconnection mechanism at the base of the Solar Corona, as we will explain in details in the following..
On the other hand we point out the presence of a TD two days later the velocity knee, on August the $8^{th}$ at $14^h34^m$, easily recognizable in Figure \ref{fig02} by looking at the switchback in the $B_x$ component and highlighted in Figure \ref{figura8}. 
 % FIGURE 8
In the latter, panels on the left hand side show (from top to bottom): magnetic field (panels a , b and c) and velocity components (panels d, e and f), and total pressure (panel g).  Magnetic field and velocity components are rotated in the magnetic field minimum variance reference system, as better shown in the 3D plots in the right hand side panels. A TD discontinuity is an MHD structure for which the total pressure is conserved, as shown in Figure \ref{figura8} panel (g), the component of the magnetic field normal to the discontinuity plane is identically zero on both sides (Figure \ref{figura8} panel c) and number density, thermal pressure and tangential component of the magnetic field vector can be discontinuous across the layer. In this particular TD, the magnetic field rotates in less than 1 minute in the minimum variance plane shown on the right-hand-side panel and characterized by robust eigenvalues ratios: $\Lambda_1 / \Lambda_3 = 0.015 $ and  $\Lambda_2 / \Lambda_3 = 0.10 $.

%0.0964328919171103 (min), lambda 2 = 0.621127880289323 (int), lambda 3 = 6.48504767784879 (max) . 

Total pressure remains roughly constant across the discontinuity and there is no relevant mass flow across the latter, because the velocity component normal to the discontinuity plane is quite constant throughout the whole time interval of Figure \ref{figura8}, indicating that it is at rest in the plasma reference frame. Looking back at Figure \ref{fig02}, magnetic field and plasma parameters rapidly become more variable upstream the TD, approaching the slow wind.  This TD seems to mark the physical border between the  rarefaction region and the surrounding slow wind. In other words, it could be the coronal discontinuity (CD) observed by \cite{McComas98} in Ulysses data during March 1995, when Ulysses was rapidly crossing the low-latitudinal region of the solar wind towards the fast polar wind flow and modeled by \cite{SchwadronMcComas2005}. The relevant difference would be that this is the first time this particular structure is recognized between the rarefaction region of an high speed stream and the following slow solar wind.

% IONI MINORI
Minor ions components of the solar wind carry relevant information about its source region, i.e. the CHBL. This is the source of a sort of transition region between the fast coronal wind and the slow wind. As a matter of fact, minor ions parameters like freezing-in temperature and relevance of low-FIP elements \citep{Geiss95a, Geiss95b, McComas2003} monotonically increase from fast to slow wind.
At this point we searched for possible signatures in plasma composition and charge state assuming the rapid depletion of fluctuations and Alfvénicity might have been due to differences in the source regions (at the basis of the solar Corona) of the stream high speed plateau and the subsequent rarefaction region. As a matter of fact, fast and slow wind largely differ for ion composition, charge state and freezing-in temperature (\cite{Bochsler IoniMinori}) and the region in between, i.e. the rarefaction region, might be a sort of wind largely mixed with the slower wind, and this would be highlighted by the composition analysis. \cite{Geiss95a} and \cite{Phillips95}, based on Ulysses observations, found several dramatic differences in ions composition between fast and slow wind. One of the most relevant is the strong bias of the slow wind in favor of low FIP (first ionization potential) elements with respect to the fast wind. \cite{Geiss95b}, using charge state ratios for C6+/C5+ and O7+/O6+ estimated the corresponding freezing-in temperatures for these elements and found that slow wind sources at the sun are quite hotter than fast wind sources. Finally, fast and slow wind greatly differ also for the higher abundance of He++ within fast and hot wind with respect to slow and cold wind \citep{Kasper2012}. 
% FIGURE 9
\begin{figure}
	\centering
	\includegraphics[width=9cm]{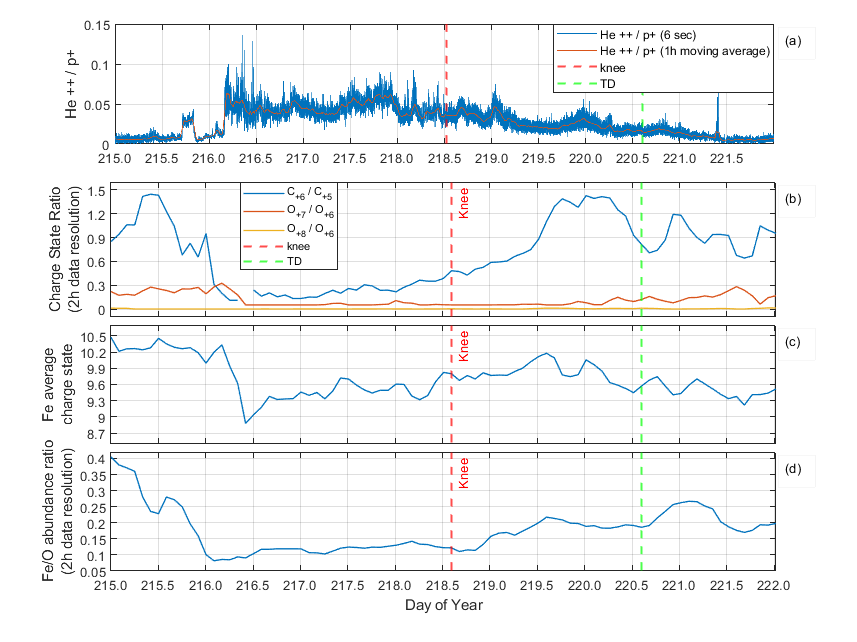}
	\caption{Temporal profiles of ions compositions during the August 2017 stream. In the top panel(a): Alpha particles over protons ratio $ {He^{++}}/{p+}$ at 6 sec time resolution (in blue) and its hourly moving average (in red). Second panel b: charge state ratio, at 2 hours time resolution, of ${C^{+6}}/{C^{+5}}$ in blue, ${O^{+7}}/{O^{+6}}$ in red and ${O^{+8}}/{O^{+6}}$ in yellow. Third panel c: iron average charge state, at 2 hours time resolution. Bottom panel d: iron over Oxygen abundance ratio, at 2 hours time resolution. The red dashed line denotes the velocity knee between fast wind region and rarefaction region.The green dashed line denotes the location of the tangential discontinuity (TD) founded at almost the end of the rarefaction region (see section \ref{sec:data}). }
	\label{figura9}
\end{figure}
The four panels of Figure \ref{figura9} show, from top to bottom, the relative abundance of He++/p+ (panel a), some charge state ratios for carbon and oxygen (panel b), charge state for iron (panel c) and relative abundance of iron over oxygen (panel d) respectively. This figure (\ref{figura9}) refers to the entire stream, i.e. the compressive region, followed by the high-speed plateau, followed by the rarefaction region and finally by the slow wind (for comparison keep the Figure \ref{fig02} in mind). As expected, the last three panels of Figure \ref{figura9} show a gradual increase from the fast wind plateau to the rarefaction region, but without a sharp jump across the velocity knee. 
% FIGURE 10
\begin{figure}
	\centering
	\includegraphics[width=9cm]{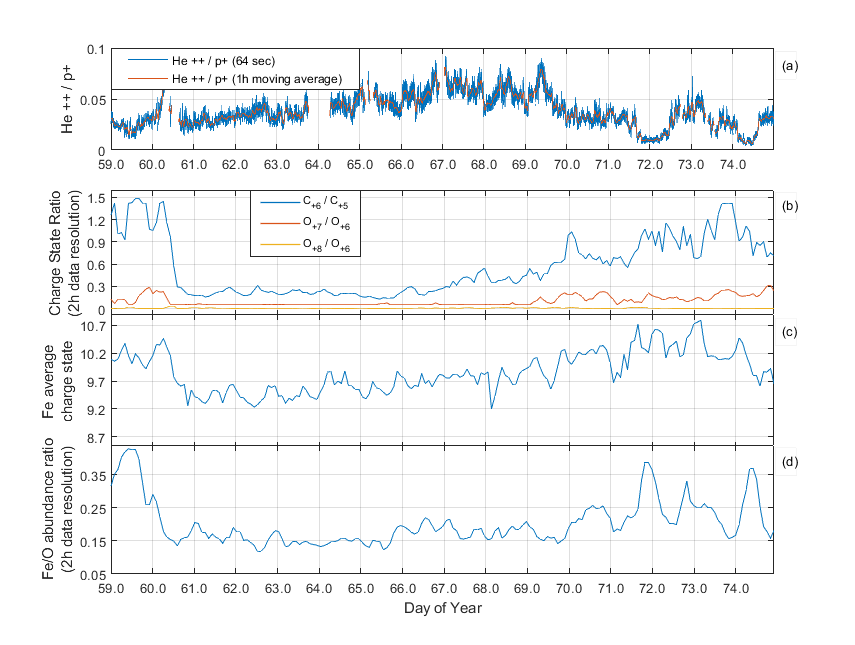}
	\caption{Temporal profiles of ions compositions during the March 2017 stream. In the top panel(a): Alpha particles over protons ratio $ {He^{++}}/{p+}$ at 64 sec time resolution (in blue) and its hourly moving average (in red). Second panel(b): charge state ratio, at 2 hours time resolution, of ${C^{+6}}/{C^{+5}}$ in blue, ${O^{+7}}/{O^{+6}}$ in red and ${O^{+8}}/{O^{+6}}$ in yellow. Third panel(c): iron average charge state, at 2 hours time resolution. Bottom panel(d): iron over Oxygen abundance ratio, at 2 hours time resolution.}
	\label{figura10}
\end{figure}
A very similar result can be seen in Figure \ref{figura10} that is the equivalent result of minor ions analysis of Figure \ref{figura9}, but referred to entire March 2017 stream. Looking at the minor ions analysis (both Figure \ref{figura9} and \ref{figura10}) we can clearly see that there are no jumps in the quantities in correspondence of the end of the fast wind plateau, nor are there large differences between the two streams of August and March.
As a matter of fact, although the time resolution of minor ions parameters provided by ACE related to the selected time period is not very high to allow to estimate small scale structures, but it's enough to show that there are not strong discontinuities in ions compositions at the location of the velocity knee such to justify, in terms of differences in the source region, abrupt changes in the fluctuations. The top panel shows the He++/p+ ratio and, although higher values for this parameter are found within the high speed plateau, the transition across the velocity knee is quite smooth. Thus, the results from this composition analysis point towards a smooth transition from fast to slow wind and did not unravel dramatic changes to be associated with the observed fluctuations abrupt depletion. Therefore, these similarities strongly suggest that the reason for the sudden depletion of turbulence observed within the stream of August 2017 and not observed within the stream of March 2017 can not be hidden in differences in the source region at photospheric level.
Moreover, the charge state ratio profile of $C_{+6}/C_{+5}$ (see the blue curve on panel b of both Figures \ref{figura9} and \ref{figura10}) is in agreement with the results shown by \cite{Schwadron2005}. They modeled a solar minimum configuration that gives rise to a co-rotating region considering the differential motion of magnetic field foot points at the Sun across the CHBL. In their model, they started the simulation at $30 R_s$ ($ \sim 0.14$  AU ) with the presence of a discontinuity at the stream interface (that comes before the fast wind plateau) and another discontinuity at the CHBL. They showed that the discontinuity at the stream interface remains whereas the discontinuity at the CHBL, in the rarefaction region, is eroded as the distance from the Sun increases (they have extended the simulation up to 5 AU). This result is in agreement with what we observe in the carbon charge ratio (panel b of Figure \ref{figura9} and Figure \ref{figura10}): we can notice a rapid decrease at the stream interface, corresponding to CH discontinuity before the fast wind plateau and, more important, we also observe a gradual increase of $C_{+6}/C_{+5}$  in the rarefaction region, corresponding to the CHBL described in model of \cite{Schwadron2005}.
% FIGURE 11
\begin{figure}
	\centering
	\includegraphics[width=9.2cm]{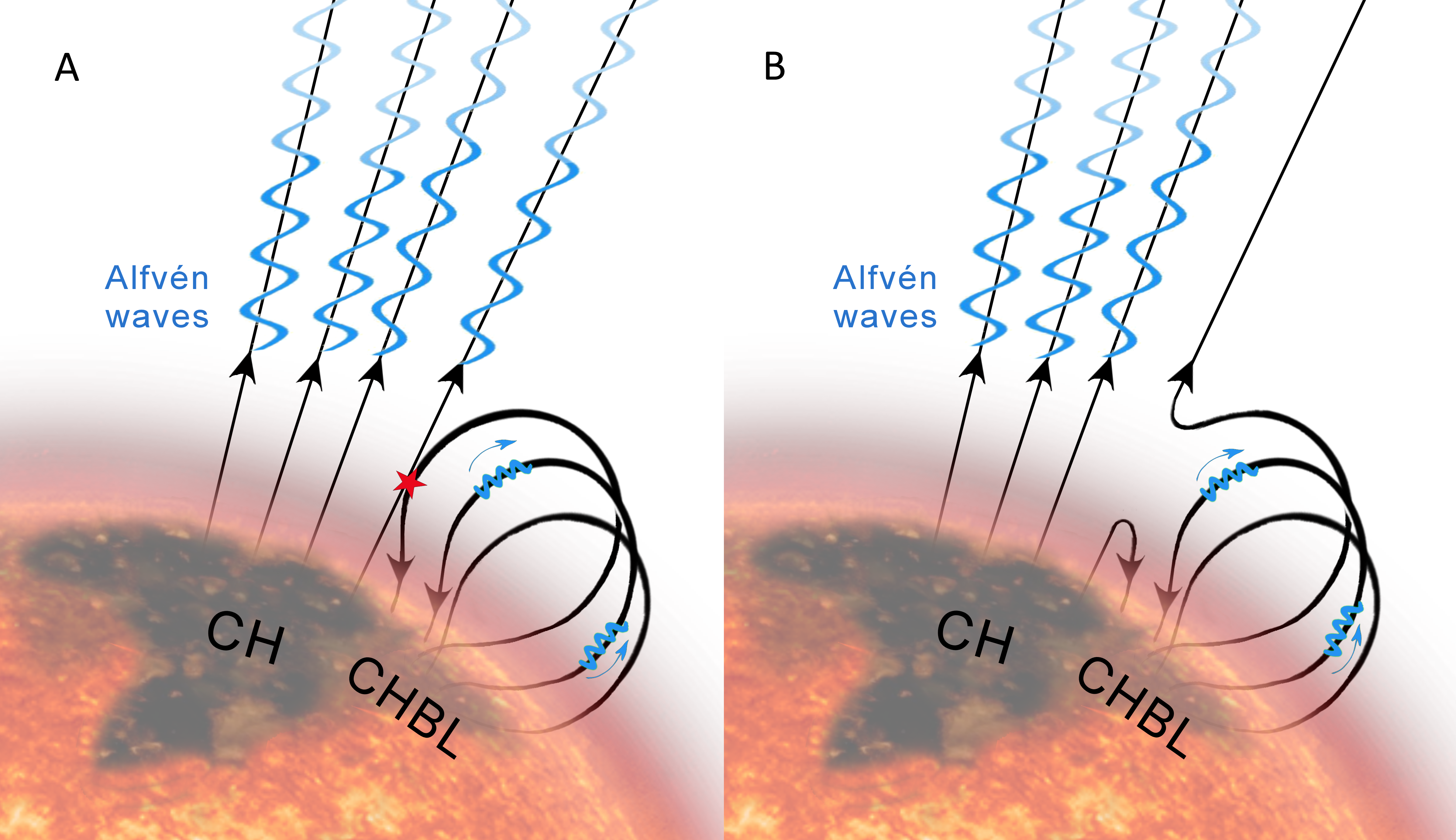}
	\caption{Proposed mechanism of interchange reconnection invoked to explain the abrupt depletion of Alfvénic fluctuations often observed in co-rotating high velocity streams. More details in the text.}
	\label{figura11}
\end{figure}

At this point we invoke a simple mechanism, as shown in Figure \ref{figura11}, which might solve this puzzle, taking into account that there is no TD there, nor is there an abrupt change in the composition analysis of the SW. The left-hand panel (Figure \ref{figura11} A) sketches open magnetic field lines rooted inside a coronal hole, indicated by CH, and coronal loops rooted inside a coronal hole boundary layer, indicated by CHBL, encircling the CH. Field lines inside the CH are populated by Alfv\'enic fluctuations while coronal loops are not. 
Although it is expected that Alfvénic fluctuations largely populate the photospheric magnetic carpet \citep{Title1998} transverse Alfvénic fluctuations would experience  a different fate in open or closed magnetic structures since in the latter ones the presence of counter-propagating Alfvénic fluctuations would ignite turbulence process with consequent transfer of energy to smaller and smaller scales to eventually dissipate and heat the plasma. Thus, we would expect smaller amplitude and weaker Alfvénic character in those fluctuations coming from originally closed field regions and larger amplitude and stronger Alfvénic character for those coming from open-field-line regions.
Interchange reconnection between open and closed field lines plays a major role in the regions bordering the coronal hole \citep{Fisk99, Fisk2001, Fisk2005, Fisk2020,Schwadron2005}. This mechanism governs the diffusion of open magnetic flux outside the CH. The motion of foot-points of magnetic field lines due to photospheric dynamics is at the basis of interchange reconnection, which tends to mix open field line regions with nearby closed loop regions. Within this framework, an open field line region, populated by propagating Alfv\'enic fluctuations, might reconnect with a closed loop region and cause a sudden disappearance of these fluctuations as shown in the right-hand panel (Figure \ref{figura11} B). This phenomenon might happen at the base of the corona during the initial phase of the wind expansion \citep{Fisk2005}.

\section{Summary}

Co-rotating high speed streams are generally characterized by a high speed plateau followed by a slow velocity decrease, commonly identified as the stream rarefaction region. These two portions of the stream, in some cases, are separated by a clear velocity knee. In these cases, the knee marks also the point beyond which Alfv\'enic fluctuations, which generally populate the high speed plateau, are dramatically and rapidly depleted.  Interesting enough, this thin border region is not characterized by any relevant TD that could be responsible for such an abrupt depletion of the fluctuations. On the other hand, we could not find either clue in the composition analysis which might have indicated a clear different origin in the source region between the wind observed in the high speed plateau and that observed in the rarefaction region. Thus, the origin of the observed rapid depletion of the fluctuations does not seem to be related neither to differences at the source region nor to a TD separating the high speed plateau from the following rarefaction region. 
Not excluding the hypothesis that this reduction may occur during the expansion of the solar wind, the rapid time scales over which this abrupt decrease in fluctuations  (few minutes) and Alfvénicity (less than one hour) is observed make us lean towards the hypothesis that a magnetic reconnection event during the initial phase of the wind expansion, within the lowest layers of the solar atmosphere, might be the mechanism responsible for what we observed and described in this paper. Phenomena of interchange magnetic reconnection favored by the motion of the foot-points of magnetic field lines due to photospheric dynamics would mix together field lines originating from open field line regions, robustly populated by large amplitude Alfv\'enic fluctuations, with magnetic loops, characteristic of surrounding closed field line regions, populated by smaller amplitude and less Alfv\'enic fluctuations. 
This suggested mechanism, which we cannot prove but propose as an hypothesis at the present, will be tested in the near future as soon as the nominal phase of the Solar Orbiter mission will start and it will become possible to directly link in-situ and remote observations.  This sort of behavior could be related to the topology of the heliospheric current sheet at the moment of the crossing by the observer. Highly inclined current sheet crossings seem to result in an abrupt depletion of the fluctuations whereas during flat current sheet crossings we observe neither a clear velocity knee nor an abrupt reduction of the fluctuations.

\begin{acknowledgements}
This work was partially supported by the Italian Space Agency (ASI) under contract ACCORDO ATTUATIVO n. 2018-30-HH.O. Results from the data analysis presented in this paper are directly available from the authors. WIND data have been accessed through the NASA SPDF-Coordinated Data Analysis Web. G.C. acknowledges support from the Laboratoire de Mécanique des Fluides et d'Acoustique at the  École centrale de Lyon, France. G.C. acknowledges SWICo's recognition through the award "Premio Mariani" 2021.
\end{acknowledgements}

\end{document}